\documentclass[aps,prd,amsmath,twocolumn,amssymbaps,showpacs]{revtex4-2}
\usepackage{graphicx}  
\usepackage{dcolumn}   
\usepackage{bm}        
\usepackage{amssymb}   
\usepackage{amsmath}   
\usepackage{bbm}
\usepackage{draftcopy}
\usepackage{relsize} 
\usepackage{xfrac}    
\usepackage{slashed}
\usepackage{color}
\usepackage{footnote}
\definecolor{dgreen}{cmyk}{1.,0.,1.,0.2}        
\definecolor{orange}{cmyk}{0.,0.353,1.,0.}    

\usepackage[bookmarks]{hyperref}

\newcommand{\di}{{\rm d}}

\newcommand{\be}{\begin{equation}}
\newcommand{\ee}{\end{equation}}                                                                               
\newcommand{\bea}{\begin{eqnarray}}
\newcommand{\eea}{\end{eqnarray}}

\begin{document}
\title{Heavy-flavor mesons in a strong electric field}

\author{Jiayun Xiang and Gaoqing Cao}
\affiliation{School of Physics and Astronomy, Sun Yat-sen University, Zhuhai 519088, China}

\date{\today}

\begin{abstract}
Very strong electromagnetic field can be generated in peripheral relativistic heavy ion collisions. This work is devoted to exploring the interplay between the effects of a constant external electric field and confining potential on heavy-flavor mesons. As the corresponding vector potential linearly depends on one spatial coordinate for a constant electric field, it might be able to overcome the linear confining potential of quantum chromodynamics (QCD) and induce deconfinement. To perform analytic calculations and for comparison, one and two dimensional systems are studied together with the realistic three dimensional systems. The one dimensional Schr$\ddot{\text o}$dinger equation can be solved analytically with the help of Airy functions, and deconfinement is indeed realized when the electric field is larger than the string tension. Focus on the confining case, the two and three dimensional Schr$\ddot{\text o}$dinger equations can be solved analytically in large $r$ limit with the help of elliptic cosine/sine functions, and the wave functions are dominated by the region antiparallel to the electric field. When a more realistic potential is applied, a non-monotonic feature is found for $\Upsilon(2S)$ and $\Upsilon(3S)$-like mesons with increasing electric field.
\end{abstract}


\maketitle

\section{Introduction}
In 2020, the physicists from the University of California, Santa Barbara listed the 10 most difficult problems in modern physics. One of the problems is related to confinement in QCD. Up to now, the most important evidence of confinement comes from the first principle lattice QCD simulations~\cite{Eichten:1974af}: a linear potential between static quarks. Such a confining potential together with Coulomb potential is usually called Cornell potential~\cite{Eichten:1974af} and was then successfully applied to heavy-flavor hadrons~\cite{Rothkopf:2019ipj,Zhao:2020jqu}. The slope of the confining potential, string tension, decreases with increasing temperature, eventually deconfinement would be realized when it becomes zero~\cite{Petreczky:2010yn}. Deconfinement transition can also be characterized by the approximate order parameter, Polyakov loop, in $2+1$ flavor QCD system~\cite{Fukushima:2017csk}. It is very interesting that thus defined deconfinement was found to occur at a crossover temperature consistent with that for chiral symmetry restoration~\cite{Karsch:1998qj,Ding:2015ona}, though they seem to be two independent aspects of QCD.

Actually, there is already a very well-known linear potential in classical physics, that is, the vector potential for a constant electric field. But it is rather different from the confining potential: First, the former is anisotropic while the latter is isotropic; second, the former tends to split hadrons apart while the latter tends to bind hadrons tightly. Usually, the effect of electric field is not important in QCD systems, but the generated electromagnetic (EM) field is so large in peripheral heavy ion collisions (HICs) that the effect should not be neglected~\cite{Voronyuk:2011jd,Bzdak:2011yy,Deng:2012pc}. Mainly, three relevant topics have been covered in high energy nuclear physics: chiral anomalous transport~\cite{Huang:2013iia,Gao:2019zhk,Cao:2021jjy,Yamamoto:2021gts}, phase transition~\cite{Cao:2015cka,Wang:2017pje,Wang:2018gmj,Cao:2019hku,Cao:2020pjq,Cao:2020pjq,Cao:2023yba,Cao:2021rwx,Yamamoto:2012bd,Cao:2015dya,Endrodi:2023wwf}, and photoproduction of dileptons and hadrons in ultra-peripheral HICs~\cite{ALICE:2014eof,ALICE:2023kgv,Xing:2020hwh,Wang:2022gkd,Wang:2021kxm}. Especially, for chiral anomalous transport, chiral electric separation effect was proposed in analog to chiral magnetic effect several years ago~\cite{Huang:2013iia} and chiral electric vortical effect was recently discovered~\cite{Gao:2019zhk,Cao:2021jjy,Yamamoto:2021gts}. For phase transition, we found neutral pseudo-scalar meson condensations in parallel EM field~\cite{Cao:2015cka,Wang:2017pje,Wang:2018gmj,Cao:2019hku,Cao:2020pjq,Cao:2020pjq,Cao:2023yba,Cao:2021rwx}, and chiral symmetry and deconfinement were also explored for $2(+1)$ flavor QCD systems in a pure electric field~\cite{Yamamoto:2012bd,Cao:2015dya,Endrodi:2023wwf}. 

It is interesting to study the interplay between the linear confining and electric potentials for heavy-flavor mesons by simply applying Schr$\ddot{\text o}$dinger equation. We expect that the heavy-flavor mesons would become unstable beyond the point where the effectively felt electric field equals the string tension and deconfinemnt could happen at a larger electric field. However, the deconfinement caused by electric field is rather different from that caused by temperature: The former is a direct effect of electric field~\cite{Yamamoto:2012bd}, but the latter is an indirect effect of temperature which modifies the string tension implicitly~\cite{Petreczky:2010yn}. Previously, magnetic effect on the meson spectra had been explored by adopting a minimal coupling~\cite{Chen:2020xsr,Zhao:2020jqu}, and it is interesting that magnetic field would enhance $J/\psi^0$ mass but reduce $\eta_c$ mass. Actually, an orthogonal electric field had also been considered in Ref.~\cite{Chen:2020xsr} together with magnetic field, but that only corresponds to the pure magnetic case in some special reference since the electric field is smaller. The effect of pure electric field had been explored with the help of spherical harmonic expansions in Ref.~\cite{Biaogang:2019hij}, and the mixing between total angular momentum eigenstates and the stability has been studied in detail.

The paper tends to develop an intuition about the electric effect on heavy-flavor mesons with quantum mechanics, and solve the Schr$\ddot{\text o}$dinger equation analytically and numerically without spherical harmonic expansions. For that purpose, it is organized as follows: In Sec.~\ref{EF}, we present the main formalism for the study, and solve the Schr$\ddot{\text o}$dinger equation or one dimensional case in Sec.~\ref{1D}, two dimensional case in Sec.~\ref{2D} and three dimensional case in Sec.~\ref{3D}, respectively. A summary is given in Sec.~\ref{summary}.

\section{The effect of a constant electric field}\label{EF}
In this section, we try to study the effect of a constant electric field on heavy-flavor mesons, composed of charm or bottom (anti-)quarks, by solving the two-body Schr$\ddot{\text o}$dinger equation. For a general $d$ dimensional case, the mesonic wave function $\Psi({\bf r}_1,{\bf r}_2)$ satisfies the following static Schr$\ddot{\text o}$dinger equation:
\bea
\left[\sum_{i=1}^{d}\!\left({\hat{p}_{1i}^2\over2m_1}\!+\!{\hat{p}_{2i}^2\over2m_2}\right)\!+\!\sigma |{\bf r}_1\!-\!{\bf r}_2|\!-\!q_1\varepsilon x_1\!-\!q_2\varepsilon x_2\!-\!E\right]\!\Psi\!=\!0.\nonumber\\
\eea
Here, ${\bf r}_1$ and ${\bf r}_2$ denote the locations of quark and antiquark components of the heavy-flavor meson, respectively; $\hat{p}_{ji}\equiv-\hbar{\partial\over\partial r_{ji} }(j=1,2)$ are the corresponding momentum operators with $r_{j1}=x_j,  r_{j2}\equiv y_{j},$ and $r_{j3}=z_j$; and $q_1$ and $q_2$ are the charges of quark and antiquark, respectively. Since long-range interparticle interaction is more important in a strong electric field, the Coulomb part is simply neglected from the realistic Cornell potential and only the linear confining part is left. The string tension takes the value $\sigma=0.2~{\rm GeV}^2$ according to lattice QCD simulation~\cite{Eichten:1974af}, and the direction of electric field $\boldsymbol{\varepsilon}$ is set to be along $x$-axis without loss of generality. Of course, $E$ is the total energy of the heavy-flavor meson. 

Usually, the mesonic dynamics is composed of global and relative ones, which can be well separated by redefining the independent coordinates: ${\bf R}\equiv {m_1{\bf r}_1+m_2{\bf r}_2\over m_1+m_2}$ and ${\bf r}\equiv {{\bf r}_1-{\bf r}_2}$. Then, the Schr$\ddot{\text o}$dinger equation becomes
\bea
\!\!\!\!\!\!\!\!\left(\sum_{i=1}^{d}\!{\hat{p}_{i}^2\over2m}+\sigma\,r-q\varepsilon x\right)\Psi=\left(E-\sum_{i=1}^{d}{\hat{P}_{i}^2\over2M}+Q\varepsilon X\right)\Psi.
\eea
On the left-hand side, the quantities correspond to relative motion with the definitions: the reduced mass $m={m_1m_2\over m_1+m_2}$, the reduced charge $q={m\over m_1}q_1-{m\over m_2}q_2$, and the momentum operators $\hat{p}_i\equiv-\hbar{\partial\over\partial r_{i}}$. On the right-hand side, the quantities correspond to global motion with the definitions: the total mass $M=m_1+m_2$, the total charge $Q=q_1+q_2$, and the momentum operators $\hat{P}_i\equiv-\hbar{\partial\over\partial R_{i}}$. So, if we take variable separation for the wave function, that is, $\Psi({\bf r}_1,{\bf r}_2)=\psi({\bf r})\Phi({\bf R})$, the Schr$\ddot{\text o}$dinger equation can be reduced to two independent static equations:
\bea
\left(\sum_{i=1}^{d}{\hat{P}_{i}^2\over2M}-Q\varepsilon X\right)\Phi&=&E_g\Phi,\\
\left(\sum_{i=1}^{d}\!{\hat{p}_{i}^2\over2m}+\sigma\,r-q\varepsilon x\right)\psi&=&E_r\psi,\label{RSE}
\eea
and the total energy is given by $E=E_r+E_g$. We are mostly interested in the relative motion that is relevant to the meson spectrum. In the following, we will focus on the relative Schr$\ddot{\text o}$dinger equation \eqref{RSE}, and one, two and three dimensional cases will be explored one by one in the subsections. 

For simplicity, we assume $q<0$, then the wave function for $q>0$ can be obtained from that for $-|q|$ by taking the spatial reflection: $x\rightarrow-x$. The masses and charges of charm and bottom quarks are, respectively
 \bea
&& m_c=1.29~{\rm GeV},\ q_c={2\over3}e;\\
&& m_b=4.7~{\rm GeV},\ \ \ q_b=-{1\over3}e;
 \eea
then the reduced masses and charges of different mesons can be easily calculated, see Table.~\ref{q}. 
\renewcommand\arraystretch{1.5}
\begin{table}[!h]
\centering
\caption{The reduced masses and charges}\label{q}
\begin{tabular}{|m{1.5cm}<{\centering}|m{1.5cm}<{\centering}|m{1.5cm}<{\centering}|m{1.5cm}<{\centering}|m{1.5cm}<{\centering}|}
\hline
 \bf{Meson}& $\bf{c\bar{c}}$ & $\bf{c\bar{b}}$ & $\bf{b\bar{c}}$& $\bf{b\bar{b}}$\\
  \hline
$\bf{m/{\rm \bf{GeV}}}$ & $0.645$ & $1.012$ & $1.012$ & $2.35$ \\
 \hline
$\bf{q/e}$ & ${2\over3}$ & $0.45$ & $-0.45$ & $-{1\over3}$ \\
\hline
\end{tabular}
\end{table}
To show the relevance of our study, we compare the corresponding relative energies solved from Eq.\eqref{RSE} to experimental results in Table.~\ref{mass} for the case $\varepsilon=0$. As we can see, the confining potential indeed dominates the contributions and is more important for heavier mesons and higher excitation states.
\begin{table}[!h]
\centering
\caption{The relative energies from experiments and Eq.\eqref{RSE} }\label{mass}
\begin{tabular}{|m{2cm}<{\centering}|m{1.5cm}<{\centering}|m{1.5cm}<{\centering}|m{1.5cm}<{\centering}|m{1.5cm}<{\centering}|}
\hline
 \bf{State} & $\boldsymbol{\psi(2S)}$ &  $\boldsymbol{B_{\rm C}(2S)}$ & $\boldsymbol{\Upsilon(2S)}$ & $\boldsymbol{\Upsilon(3S)}$\\
  \hline
$\bf{E_{\rm r}^{Exp}} ({\rm GeV})$ & $1.106$ & $0.881$ & $0.623$ & $0.955$ \\
 \hline
$\bf{E_{\rm r}^{Th}}\ ({\rm GeV})$ & $0.735$ & $0.632$ & $0.477$ & $0.835$\\
\hline
\end{tabular}
\end{table}

\subsection{One dimension}\label{1D}
In the one dimensional case $d=1$, the relative Schr$\ddot{\text o}$dinger equation \eqref{RSE} becomes
\bea
\left({\hat{p}_x^2\over2m}+\sigma\,|x|+|q|\varepsilon x\right)\psi(x)=E_r\psi(x),
\eea
which can be divided into two piecewise equations
\bea
    \begin{cases}
   \left[-\frac{\hbar^2}{2m}\frac{\partial^2}{\partial x^2}-(\sigma-|q|\varepsilon)x\right]\psi_-(x)=E_r\psi_-(x), \ x<0;\\
    \\
     \left[-\frac{\hbar^2}{2m}\frac{\partial^2}{\partial x^2}+(\sigma+|q|\varepsilon)x\right]\psi_+(x)=E_r\psi_+(x), \ x>0.
    \end{cases}
\eea
Here, the electric field $\varepsilon$ induces splitting to the string tension, that is, $\sigma$ is effectively reduced to $\sigma_-\equiv\sigma-|q|\varepsilon$ for $x<0$ and enhanced to $\sigma_+\equiv\sigma+|q|\varepsilon$ for $x>0$. It is well known that the solution to the pure electric field case is given by Airy functions~\cite{Landau2008}, then the analytic solutions to the above differential equations can be obtained by modifying the variables as
\bea
 \begin{cases}
    \psi_-=C_{1-}Ai\left(a_-\left(x\!+\!{E_r\over\sigma_-}\right)\right)+C_{2-}Bi\left(a_-\left(x\!+\!{E_r\over\sigma_-}\right)\right)\\
    \\
     \psi_+=C_{1+}Ai\left(a_+\left(x\!-\!{E_r\over\sigma_+}\right)\right)+C_{2+}Bi\left(a_+\left(x\!-\!{E_r\over\sigma_+}\right)\right)
     \end{cases}
\eea
with $a_\pm\equiv\pm(\frac{2m\sigma_\pm}{\hbar^2})^{1\over3}$, and $Ai(u)$ and $Bi(u)$ two independent Airy functions. Since $a_+$ is positive definite for $|q|\varepsilon>0$, the convergence $\lim_{x\rightarrow\infty}\psi_+(x)=0$ requires $C_{2+}=0$ as in the case with pure electric field. For the piece with $x<0$, the situation is a little complicated: If $\sigma_->0$ hence $a_-<0$, we must require $C_{2-}=0$ in order that $\lim_{x\rightarrow-\infty}\psi_-(x)=0$; if $\sigma_-<0$, both $C_{1-}$ and $C_{2-}$ can be nonzero. As will be shown, the former corresponds to full confinement, while the latter corresponds to deconfinement and a boundary is needed at $-b\ (b>0)$ to bind the mesons. To completely fix the coefficients $C_{i\pm}\ (i=1,2)$ and  the relative energy $E_r$, we must refer to smooth conditions and normalization of the wave function, that is,
\bea
&&\psi_-(0)=\psi_+(0),\ \psi'_-(0)=\psi'_+(0);\\
&&\int_{x<0}| \psi_-(x)|^2\di x+\int_{x>0}| \psi_+(x)|^2\di x=1.
\eea

\subsubsection{The case $\sigma>|q|\varepsilon$}
For $\sigma_->0$, the smooth conditions imply
\bea
    \begin{cases}
        C_{1-}Ai\left(a_-{E_r\over\sigma_-}\right)=C_{1+}Ai\left(-a_+{E_r\over\sigma_+}\right),\\
        C_{1-}a_-Ai'\left(a_-{E_r\over\sigma_-}\right)=C_{1+}a_+Ai'\left(-a_+{E_r\over\sigma_+}\right),
        \end{cases}
\eea
and it follows that the relative energy $E_r$ satisfies
\bea
{Ai\left(a_-{E_r\over\sigma_-}\right)\over a_-Ai'\left(a_-{E_r\over\sigma_-}\right)}={Ai\left(-a_+{E_r\over\sigma_+}\right)\over a_+ Ai'\left(-a_+{E_r\over\sigma_+}\right)}.
 \eea
 Since Airy functions satisfy the integral property
\bea
\int f(u)g(u)\di u=u f(u)g(u)-f'(u)g'(u)+C\label{int}
\eea
for $f(u),g(u)$ being any of $Ai(u), Bi(u)$, the normalization condition can be translated to
\bea
1&=&|C_{1+}|^2\left[{E_r\over\sigma_+}Ai^2\left(-a_+{E_r\over\sigma_+}\right)+{1\over a_+}Ai'^2\left(-a_+{E_r\over\sigma_+}\right)\right]\nonumber\\
&+&|C_{1-}|^2\left[{E_r\over\sigma_-}Ai^2\left(a_-{E_r\over\sigma_-}\right)- {1\over a_-}Ai'^2\left(a_-{E_r\over\sigma_-}\right)\right].
\eea
In the limit $\varepsilon\rightarrow0$, the smooth conditions would be reduced to
\bea
C_{1-}Ai\left(-a{E_r\over\sigma}\right)=C_{1+}Ai\left(-a{E_r\over\sigma}\right),\\
-C_{1-}Ai'\left(-a{E_r\over\sigma}\right)=C_{1+}Ai'\left(-a{E_r\over\sigma}\right),
 \eea
with $a\equiv(\frac{2m\sigma}{\hbar^2})^{1\over3}$. And it follows that $Ai\left(-a{E_r\over\sigma}\right)=0, C_{1-}=- C_{1+}$ or $Ai'\left(-a{E_r\over\sigma}\right)=0, C_{1-}=C_{1+}$, which correspond to odd and even wave functions, respectively. Then, the normalization condition fixes the coefficients to
 \bea
&&C_{1-}=- C_{1+}=\pm\sqrt{a\over 2}Ai'^{-1}\left(-a{E_r\over\sigma}\right)
\eea
or
 \bea
&&C_{1-}= C_{1+}=\pm\sqrt{\sigma\over 2E_r}Ai^{-1}\left(-a{E_r\over\sigma}\right).
\eea
\begin{figure}[!htb]
	\begin{center}
	\includegraphics[width=8cm]{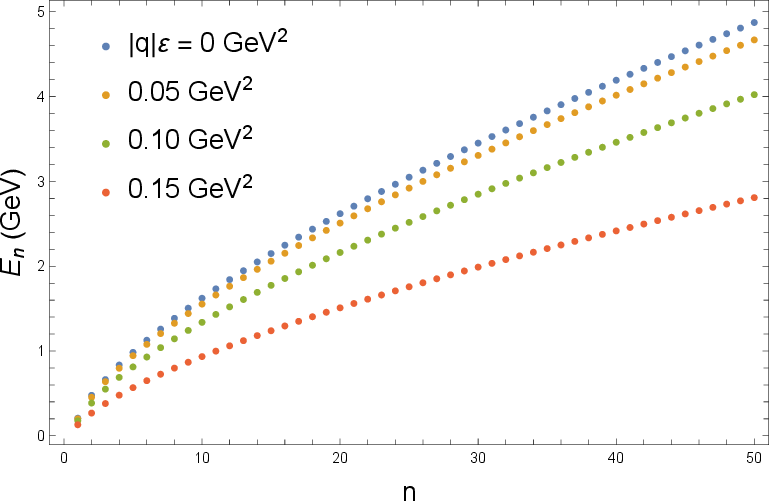}
	\caption{The lowest eigenenergies $E_{n}$ of bottomium with $n=1,2,\dots,50$ for different supercritical electric fields.}\label{Enn1}
	\end{center}
\end{figure}
\begin{figure}[!htb]
	\begin{center}
	\includegraphics[width=8cm]{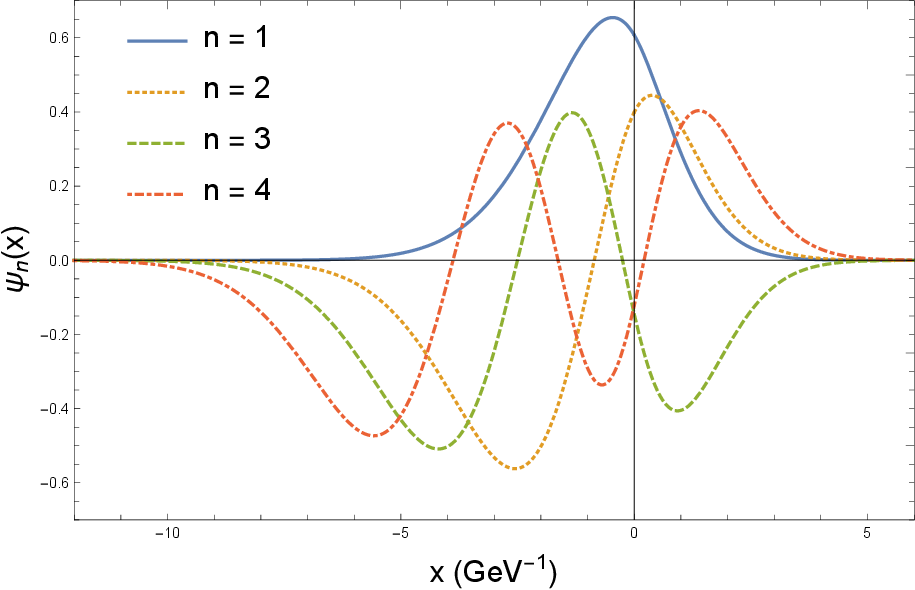}
	\caption{The lowest four eigenfunctions of bottomium for the supercritical electric field $|q|\varepsilon=0.1~{\rm GeV}^2$.}\label{psinn1}
	\end{center}
\end{figure}

Take bottomium for example, the lowest $50$ eigenenergies $E_{n}$ are listed in Fig.~\ref{Enn1} for different subcritical electric fields. As we can see, the eigenenergies get reduced by the external electric field since the dominated effective string tension $\sigma_-$ decreases. For demonstration, we show the lowest four eigenfunctions in Fig.~\ref{psinn1} for the electric field $|q|\varepsilon=0.1~{\rm GeV}^2$. It is found that the number of extrema is the same as the order $n$ of the eigenenergy $E_{n}$ and the wavefunction is more dispersed for larger $n$. Actually, compared to the case with $|q|\varepsilon=0$, the strongest extrema shift antiparallel to the electric field for all $n$, indicating splitting dynamics.

\subsubsection{The case $\sigma<|q|\varepsilon$}
For $\sigma_-<0$, the smooth conditions gives
\bea
\begin{cases}
	C_{1-}Ai\left(a_-{E_r\over\sigma_-}\right)+C_{2-}Bi\left(a_-{E_r\over\sigma_-}\right)\\\ \ \ \ =C_{1+}Ai\left(-a_+{E_r\over\sigma_+}\right),\\
	C_{1-}a_-Ai'\left(a_-{E_r\over\sigma_-}\right)+	C_{2-}a_-Bi'\left(a_-{E_r\over\sigma_-}\right)\\\ \ \ \ =C_{1+}a_+Ai'\left(-a_+{E_r\over\sigma_+}\right),
\end{cases}\label{SC}
\eea
and the boundary condition at $-b$ is
\bea
C_{1-}Ai\left(a_-\left({E_r\over\sigma_-}\!-\!b\right)\right)+C_{2-}Bi\left(a_-\left({E_r\over\sigma_-}\!-\!b\right)\right)=0.\nonumber\\\label{BC}
\eea
Then, the requirement of nontrivial solutions $C_{1-},C_{2-}$ and $C_{1+}$ to the linear equations \eqref{SC} and \eqref{BC} helps to fix the relative energy, that is,
\bea
	\!\!\left | \begin{matrix}
		Ai\left(a_-{E_r\over\sigma_-}\right) &\!\!Bi\left(a_-{E_r\over\sigma_-}\right)&\!\!\!\!-Ai\left(-a_+{E_r\over\sigma_+}\right)\\
	a_-Ai'\left(a_-{E_r\over\sigma_-}\right)&\!\!a_-Bi'\left(a_-{E_r\over\sigma_-}\right)&\!\!-a_+Ai'\left(-a_+{E_r\over\sigma_+}\right)\\
	Ai\!\left(\!a_-\!\left({E_r\over\sigma_-}\!-\!b\!\right)\!\!\right)&\!\!Bi\!\left(\!a_-\!\left({E_r\over\sigma_-}\!-\!b\!\right)\!\!\right)&\!\!\!\!0
	\end{matrix}\right|=0.\nonumber\\
\eea
By applying the integral property \eqref{int}, the normalization condition becomes 
\bea
1&=&|C_{1+}|^2\left[{E_r\over\sigma_+}Ai^2\left(-a_+{E_r\over\sigma_+}\right)+{1\over a_+}Ai'^2\left(-a_+{E_r\over\sigma_+}\right)\right]\nonumber\\
&+&\!{E_r\over\sigma_-}\left|C_{1-}Ai\left(a_-{E_r\over\sigma_-}\right)\!+\!C_{2-}Bi\left(a_-{E_r\over\sigma_-}\right)\right|^2\!\!\!-\!\left({E_r\over\sigma_-}\!-\!b\!\right)\nonumber\\
&-&\left|C_{1-}Ai\!\left(\!a_-\!\left({E_r\over\sigma_-}\!-\!b\!\right)\!\!\right)\!+\!C_{2-}Bi\!\left(\!a_-\!\left({E_r\over\sigma_-}\!-\!b\!\right)\!\!\right)\right|^2\nonumber\\
&-&{1\over a_-}\left[\left|C_{1-}Ai'\left(a_-{E_r\over\sigma_-}\right)\!+\!C_{2-}Bi'\left(a_-{E_r\over\sigma_-}\right)\right|^2-\right.\nonumber\\
&&\!\!\!\!\!\!\left.\left|C_{1-}Ai'\!\left(\!a_-\!\left({E_r\over\sigma_-}\!-\!b\!\right)\!\!\right)\!+\!C_{2-}Bi'\!\left(\!a_-\!\left({E_r\over\sigma_-}\!-\!b\!\right)\!\!\right)\right|^2\right].
\label{NC}
\eea
As $C_{1-}\propto C_{2-}\propto C_{1+}$ according to Eqs.~\eqref{SC} and \eqref{BC}, they can be easily solved from Eq.~\eqref{NC} by canceling two coefficients beforehand. 

\begin{figure}[!htb]
	\begin{center}
	\includegraphics[width=8cm]{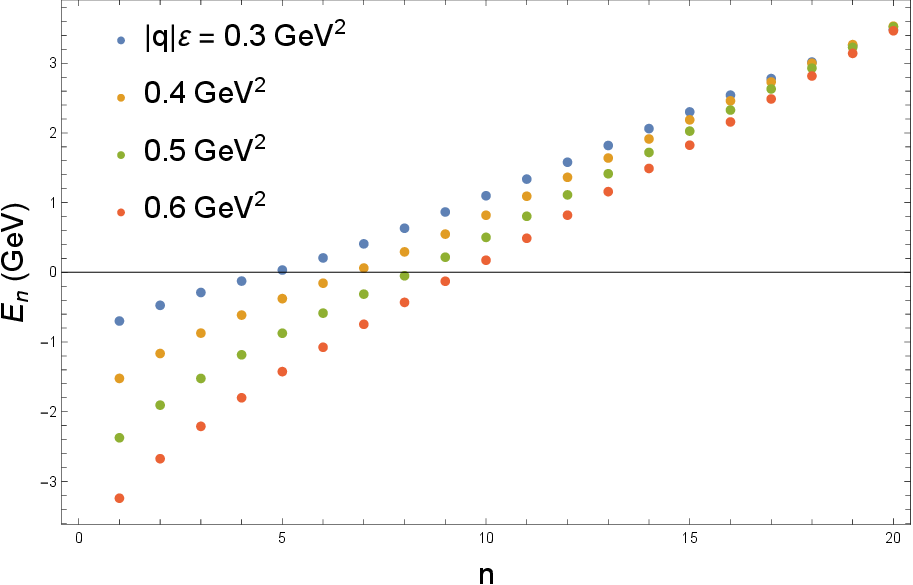}
	\caption{The lowest eigenenergies $E_{n}$ of bottomium with $n=1,2,\dots,20$ for different supercritical electric fields with $b=10~{\rm GeV}$.}\label{Enp1}
	\end{center}
\end{figure}

\begin{figure}[!htb]
	\begin{center}
	\includegraphics[width=8cm]{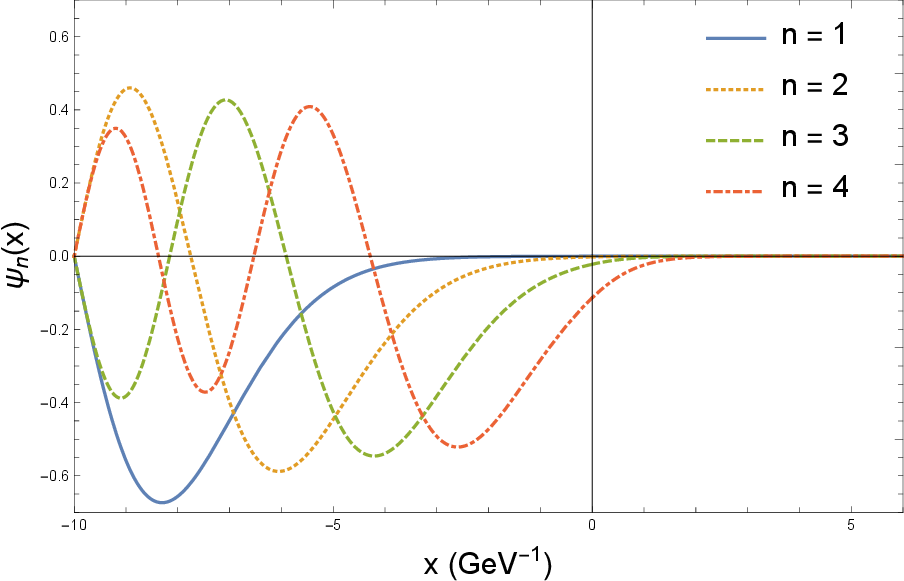}
	\caption{The lowest four eigenfunctions of bottomium for the supercritical electric field $|q|\varepsilon=0.3~{\rm GeV}^2$ with $b=10~{\rm GeV}$.}\label{psinp1}
	\end{center}
\end{figure}

The lowest $20$ eigenenergies $E_{n}$ of bottomium are listed in Fig.~\ref{Enp1} for different supercritical electric fields: Though the eigenenergy decreases with $|q|\varepsilon$ for small $n$, it mildly depends on $|q|\varepsilon$ for large $n$ and is consistent with the half free case $|q|\varepsilon=\sigma$. The lowest four eigenfunctions are negative and demonstrated in Fig.~\ref{psinp1} for $|q|\varepsilon=0.3~{\rm GeV}^2$: All the features are the same as those in Fig.~\ref{psinn1}, but they all strongly oscillate around the boundary $-b$ which actually signals deconfinement. 
\subsection{Two dimensions}\label{2D}
In the two dimensional case $d=2$, the relative Schr$\ddot{\text o}$dinger equation \eqref{RSE} becomes
\bea
\left({\hat{p}_x^2+\hat{p}_y^2\over2m}+\sigma\,r+|q|\varepsilon x\right)\psi(x,y)=E_r\psi(x,y)\label{SE2}
\eea
with $r=\sqrt{x^2+y^2}$. For a Coulomb potential and a constant electric field, the variables can be separated in the wave function by referring to the parabolic coordinate with $x\equiv{1\over2}(\tau_1^2-\tau_2^2)$ and $y\equiv\pm \tau_1\tau_2$~\cite{Landau2008}. However,  for a confining potential and a constant electric field, such a simplification no longer applies, thus we have to solve the Schr$\ddot{\text o}$dinger equation \eqref{SE2} directly by performing numerical calculations. We will take a close look at the potential in polar coordinate which becomes
\bea
\sigma\,r+|q|\varepsilon x=\left(\sigma+|q|\varepsilon \cos \theta\right) r
\eea
after the redefinitions $x\equiv r\cos \theta$ and $y\equiv r\sin \theta$. Compared to the one dimensional case, an $\theta$-dependent effective string tension is involved in the two dimensions, that is, $\sigma(\theta)\equiv\sigma+|q|\varepsilon \cos \theta$, which changes continuously from parallel to $\boldsymbol{\varepsilon}$ to antiparallel to $\boldsymbol{\varepsilon}$ with continuous change of $\theta$. And one can tell that the one in the parallel direction is the most confined and always positive, while the antiparallel one is the least confined and can be negative. Similar to the one dimensional case, $\sigma<|q|\varepsilon$ would render the wave function unbound in the antiparallel direction, thus $|q|\varepsilon_c=\sigma$ is also the critical electric field of deconfinement.

We are mostly interested in the confining case, that is, $\sigma>|q|\varepsilon$. For comparison, it is useful to discuss the case with only a confining potential first. In the limit $\varepsilon\rightarrow0$, the Schr$\ddot{\text o}$dinger equation \eqref{SE2} can be reduced to
\bea
\left[-{\hbar^2\over2m}\left({\partial^2\over\partial r^2}+{1\over r}{\partial\over\partial r}-{l^2\over r^2}\right)+\sigma\,r\right]R(r)=E_rR(r)
\eea
by defining $\psi(x,y)\equiv e^{i\,l\,\theta}R(r)$ in polar coordinate. It can be further simplified by redefining $R(r)=u(r)/\sqrt{r}$ and we have
\bea
\left[-{\hbar^2\over2m}\left({\partial^2\over\partial r^2}+{{1\over4}-l^2\over r^2}\right)+\sigma\,r\right]u(r)=E_ru(r). \label{SEr2}
\eea
For a given orbital angular momentum $l$, the $n$-th lowest eigenenergy $E_r^{nl}$ can be directly evaluated from Eq.\eqref{SEr2}, and it is found that $E_r^{nl}$ increases with $l$ for a given $n$. Actually, the large $r$ part is dominated by the confining potential for a given $E_r^{nl}$ and the Schr$\ddot{\text o}$dinger equation \eqref{SEr2} can be reduced to
\bea
\left[-{\hbar^2\over2m}{\partial^2\over\partial r^2}+\sigma\,r\right]u(r)=0.
\eea
This is effectively a one dimensional differential equation, thus the large $r$ part of $u(r)$ can be well approximated by $u(r)=CAi\left(ar\right)$ by following the discussions in Sec.\ref{1D}, regardless of $l$ or $n$. So, $\psi(x,y)\propto e^{i\,l\,\theta}Ai\left(ar\right)/\sqrt{r}$ for large $r$.

For a finite $\varepsilon$, the Schr$\ddot{\text o}$dinger equation can be transformed to 
\bea
&&\left[-{\hbar^2\over2m}\left({\partial^2\over\partial r^2}\!+\!{1\over 4r^2}\!+\!{1\over r^2}{\partial^2\over\partial \theta^2}\right)+(\sigma\!+\!|q|\varepsilon \cos \theta)\,r\right]u(r,\theta)\nonumber\\
&&=E_ru\left(r,\theta\right)
\eea
by redefining $\psi(x,y)=u\left(r,\theta\right)/\sqrt{r}$ in polar coordinate. If we assume $u\left(r,\theta\right)\equiv\Theta(r,\theta)\upsilon(r)$ and ${\partial^2\over\partial r^2}\Theta\left(r,\theta\right)$ is small, it can be separated into two coupled differential equations:
\bea
\left[-{\hbar^2\over2mr^2}{\partial^2\over\partial \theta^2}+|q|\varepsilon r \cos \theta-\epsilon(r)\right]\Theta(r,\theta)&=&0,\label{dt}\\
\left[-{\hbar^2\over2m}\left({\partial^2\over\partial r^2}\!+\!{1\over 4r^2}\right)+\sigma\,r+\epsilon(r)-E_r\right]\upsilon(r)&=&0.\label{dr}
\eea
The solutions to Eq.\eqref{dt} are Mathieu functions: $S({8mr^2\epsilon(r)\over \hbar^2},{4|q|\varepsilon mr^3\over \hbar^2},{\theta\over2})$ and $C({8mr^2\epsilon(r)\over \hbar^2},{4|q|\varepsilon mr^3\over \hbar^2},{\theta\over2})$. And the requirement of $2\pi$-periodicity constrains the eigenenergy $\epsilon_{2n}(r)$ to 
\bea
&&{\hbar^2\over8mr^2}b_{2n}\left({4|q|\varepsilon mr^3\over \hbar^2}\right)\ (n=1,2,\dots),\nonumber\\
&&{\hbar^2\over8mr^2}a_{2n}\left({4|q|\varepsilon mr^3\over \hbar^2}\right)\ (n=0,1,2,\dots),
\eea
where $b_{2n}(p)$ and $a_{2n}(p)$ are the characteristic values of $S(a,p,{\theta\over2})$ and $C(a,p,{\theta\over2})$, respectively. Accordingly, the eigenfunctions $\Theta_{2n}(r,\theta)$ are given by elliptic cosine and sine functions:
\bea
&&se_{2n}\left({\theta\over2},{4|q|\varepsilon mr^3\over \hbar^2}\right)\ (n=1,2,\dots),\nonumber\\
&&ce_{2n}\left({\theta\over2},{4|q|\varepsilon mr^3\over \hbar^2}\right)\ (n=0,1,2,\dots),
\eea
and Eq.\eqref{dr} becomes
\bea
\left[\!-{\hbar^2\over2m}\left({\partial^2\over\partial r^2}\!+\!{1\over 4r^2}\right)\!+\!\sigma\,r\!+\!\epsilon_{2n}(r)\!-\!E_r\!\right]\!\upsilon(r)=0.
\eea
In the large $p$ limit, $b_{2n}(p)\approx a_{2n}(p)\approx-2p$, hence $\epsilon_{2n}(r)\approx-|q|\varepsilon r$ in the large $r$ limit. Correspondingly, $\Theta_{2n}(r,\theta)$ oscillates in a very narrow range around $\theta=\pi$, consistent with the expectation from the classical electromagnetism. It is checked that $\Theta_{2n}(r,\theta)$ changes slowly with $r$ for given $n$ and $\theta$, so ${\partial^2\over\partial r^2}\Theta\left(r,\theta\right)$ is indeed small compared to the $r$-linear terms. Eventually, the solution follows as $\psi(x,y)\propto \Theta_{2n}(r,\theta)Ai\left(a_-r\right)/\sqrt{r}$ in the large $r$ limit. 

\begin{figure}[!htb]
	\begin{center}
	\includegraphics[width=8cm]{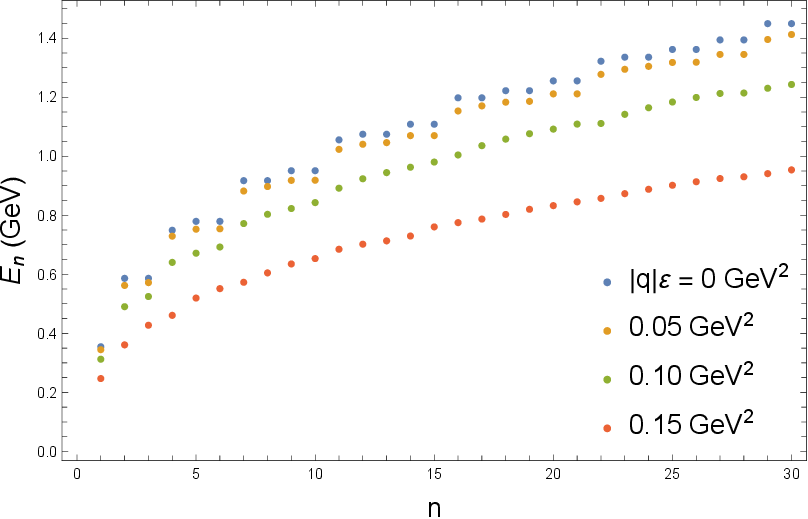}
	\caption{The lowest eigenenergies $E_{n}$ of bottomium with $n=1,2,\dots,30$ for different subcritical electric fields.}\label{Enn2}
	\end{center}
\end{figure}

\begin{figure}[!htb]
	\begin{center}
	\includegraphics[width=8cm]{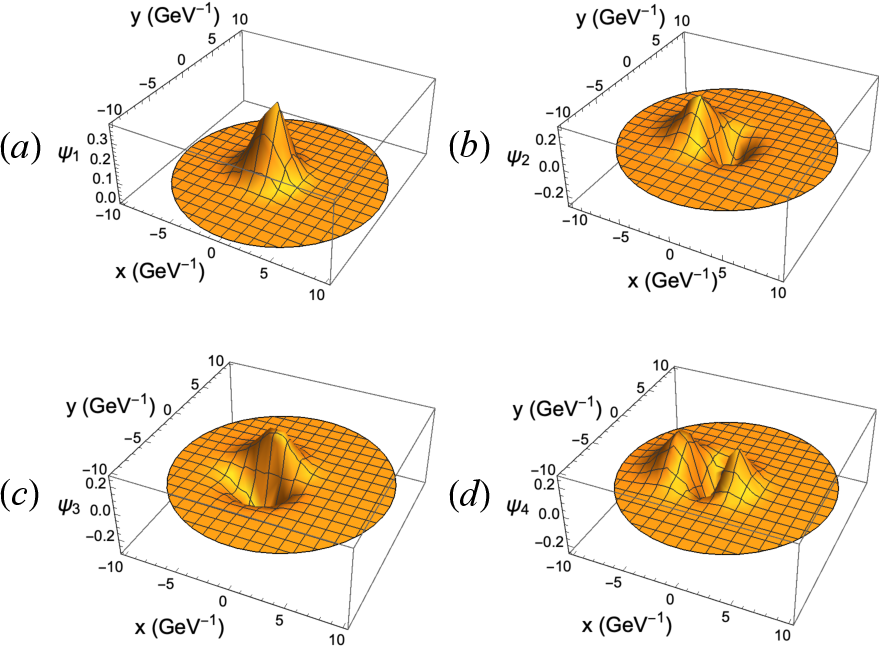}
	\caption{The lowest four eigenfunctions of bottomium for the subcritical electric field $|q|\varepsilon=0.1~{\rm GeV}^2$. $(a), (b), (c)$ and $(d)$ correspond to $\psi_{10}(x,y), \psi_{11}(x,y)\mp\psi_{1(-1)}(x,y)$ and $\psi_{20}(x,y)$ at $|q|\varepsilon=0$, respectively. }\label{psinn2}
	\end{center}
\end{figure}

The full Schr$\ddot{\text o}$dinger equation \eqref{SE2} can be solved numerically for different subcritical electric fields and the lowest $30$ eigenenergies $E_{n}$ of bottomium are listed in Fig.~\ref{Enn2}. As we can see, there is usually one or two degenerate states for a given eigenenergy when $|q|\varepsilon=0$: If we check the $l^2$-dependence of Eq.\eqref{SEr2}, it can be easily understood that the one degenerate state corresponds to $l=0$ and the two degenerate states correspond to $l=\pm1,\pm2,\dots$. Similar to that in Fig.~\ref{Enn1}, the eigenenergies get reduced by the external electric field; and the double degeneracy breaks down because the angular momentum is no longer a good quantum number in an electric field. For demonstration, we show the lowest four eigenfunctions in Fig.~\ref{psinn2} for the electric field $|q|\varepsilon=0.1~{\rm GeV}^2$: $(a), (b), (c)$ and $(d)$ correspond to $\psi_{10}(x,y), \psi_{11}(x,y)\mp\psi_{1(-1)}(x,y)$ and $\psi_{20}(x,y)$ at $|q|\varepsilon=0$, respectively. It is found that the centers of the wave functions shift antiparallel to $\boldsymbol{\varepsilon}$, consistent with the extremely anisotropic feature in the large $r$ limit.

\subsection{Three dimensions}\label{3D}
In the three dimensional case $d=3$, the relative Schr$\ddot{\text o}$dinger equation \eqref{RSE} becomes
\bea
\left({\hat{p}_x^2+\hat{p}_y^2+\hat{p}_z^2\over2m}+\sigma\,r+|q|\varepsilon x\right)\psi(x,y,z)=E_r\psi(x,y,z)\nonumber\\ \label{SE3}
\eea
with $r=\sqrt{x^2+y^2+z^2}$. Similar to the case with two dimensions, for a Coulomb potential and a constant electric field, the variables can be separated in the wave function by referring to the parabolic coordinate with $x\equiv{1\over2}(\tau_1^2-\tau_2^2), y\equiv\pm \tau_1\tau_2\cos\theta$ and $z\equiv\pm \tau_1\tau_2\sin\theta$. For a confining potential and a constant electric field, no such a simplification can apply. However, there is rotational symmetry around $x$-axis, so the Schr$\ddot{\text o}$dinger equation can be reduced to
\bea
\left[-{\hbar^2\over2m}\!\left({\partial^2\over\partial \rho^2}\!+\!{{1\over4}\!-\!l^2\over \rho^2}\!+\!{\partial^2\over\partial x^2}\right)\!+\!\sigma\,r\!+\!|q|\varepsilon x\!-\!E_r\right]\!\!u(\rho,x)=0\nonumber\\\label{SE3C}
\eea
with $r=\sqrt{\rho^2+x^2}$ by defining the wave function $\psi(x,y,z)\equiv e^{i\,l\,\phi}u(\rho,x)/\sqrt{\rho}$ in cylindrical coordinate, see the discussions in Sec.\ref{2D}. For a given orbital angular momentum $l$, this is effectively a two dimensional differential equation and can be solved numerically. Similar to the two dimensional case, a $\theta$-dependent effective string tension can be identified if we define $x\equiv r\cos \theta$ and $\rho\equiv r\sin \theta$. Then, the one antiparallel to $\boldsymbol{\varepsilon}$ is the smallest and $|q|\varepsilon_c=\sigma$ is also the critical electric field of deconfinement.

We are mostly interested in the confining case, that is, $\sigma>|q|\varepsilon$. For comparison, it is useful to discuss the case with only a confining potential first. In the limit $\varepsilon\rightarrow0$, the Schr$\ddot{\text o}$dinger equation \eqref{SE3} can be reduced to
\bea
\left[-{\hbar^2\over2m}\left({\partial^2\over\partial r^2}+{2\over r}{\partial\over\partial r}-{L(L+1)\over r^2}\right)+\sigma\,r\right]R(r)=E_rR(r)\nonumber\\
\eea
by defining $\psi(x,y,z)\equiv Y_{Ll}(\theta,\phi)R(r)$ in spherical coordinate with $Y_{Ll}(\theta,\phi)$ the spherical harmonics. It can be further simplified by redefining $R(r)=u(r)/r$ and we have
\bea
\left[-{\hbar^2\over2m}\left({\partial^2\over\partial r^2}-{L(L+1)\over r^2}\right)+\sigma\,r\right]u(r)=E_ru(r). \label{SEr3}
\eea
For a given orbital angular momentum $L$, the $n$-th lowest eigenenergy $E_r^{nL}$ can be directly evaluated from Eq.\eqref{SEr3}, and it is found that $E_r^{nL}$ increases with $L$ for a given $n$. Similar to the case with two dimensions, the large $r$ part of $u(r)$ can be well approximated by $u(r)=CAi\left(ar\right)$ by following the discussions in Sec.\ref{1D}, regardless of $L$ or $n$. So, $\psi(x,y,z)\propto Y_{Ll}(\theta,\phi)Ai\left(ar\right)/r$ for large $r$.

For a finite $\varepsilon$, the Schr$\ddot{\text o}$dinger equation \eqref{SE3C} can be transformed to 
\bea
&&\bigg[-{\hbar^2\over2m}\left({\partial^2\over\partial r^2}+{1\over r^2}{\partial^2\over\partial \theta^2}\!+\!{1\over r^2}\left({1\over4}\!+\!{{1\over4}-l^2\over \sin^2\theta}\right)\right)\nonumber\\
&&\ \ \ \ \ \ \ \ \ \ \ \ \ \ \ +(\sigma\!+\!|q|\varepsilon \cos \theta)\,r\bigg]U(r,\theta)=E_rU\left(r,\theta\right)
\eea
by redefining $u(\rho,x)\equiv U\left(r,\theta\right)/\sqrt{r}$ in spherical coordinate. Similar to the two dimensional case, it can be separated into two coupled differential equations:
\bea
\left[-{\hbar^2\over2mr^2}\left({\partial^2\over\partial \theta^2}\!+\!{{1\over4}\!-\!b^2\over \sin^2\theta}\right)\!\!+\!|q|\varepsilon r \cos \theta\!-\!\epsilon(r)\right]\Theta^l(r,\theta)&=&0,\nonumber\\\label{dt3}\\
\left[-{\hbar^2\over2m}\left({\partial^2\over\partial r^2}\!+\!{1\over 4r^2}\right)+\sigma\,r+\epsilon(r)-E_r\right]\upsilon\left(r\right)&=&0,\nonumber\\\label{dr3}
\eea
if we assume $U\left(r,\theta\right)\equiv\Theta^l(r,\theta)\upsilon(r)$ and ${\partial^2\over\partial r^2}\Theta^l\left(r,\theta\right)$ is small. Compared to Eqs.~\eqref{dt} and \eqref{dr} in two dimensions, the only difference is that there is an extra term $\propto  \sin^{-2}\theta$ in the differential equation of $\theta$ \eqref{dt3}, which then cannot be solved exactly with a special function. However,  in the large $r$ limit, such a term can be neglected and the solution are approximately Mathieu functions: $S({8mr^2\epsilon(r)\over \hbar^2},{4|q|\varepsilon mr^3\over \hbar^2},{\theta\over2})$ and $C({8mr^2\epsilon(r)\over \hbar^2},{4|q|\varepsilon mr^3\over \hbar^2},{\theta\over2})$. Then, all the discussions in two dimensions can apply in the large $r$ limit and we expect the quark-antiquark system to mainly align antiparallel to $\boldsymbol{\varepsilon}$ as well. Though 
$\sin^{-2}\theta$ diverges at $\theta=\pi$, it would not cause any problem in the wave function, since it is there for the case $|q|\varepsilon=0$ but the corresponding eigenfunctions are smooth. At most, the eigenfunction would more widely disperse around $-\boldsymbol{\varepsilon}$ compared to that of two dimensions.

\begin{figure}[!htb]
	\begin{center}
	\includegraphics[width=8cm]{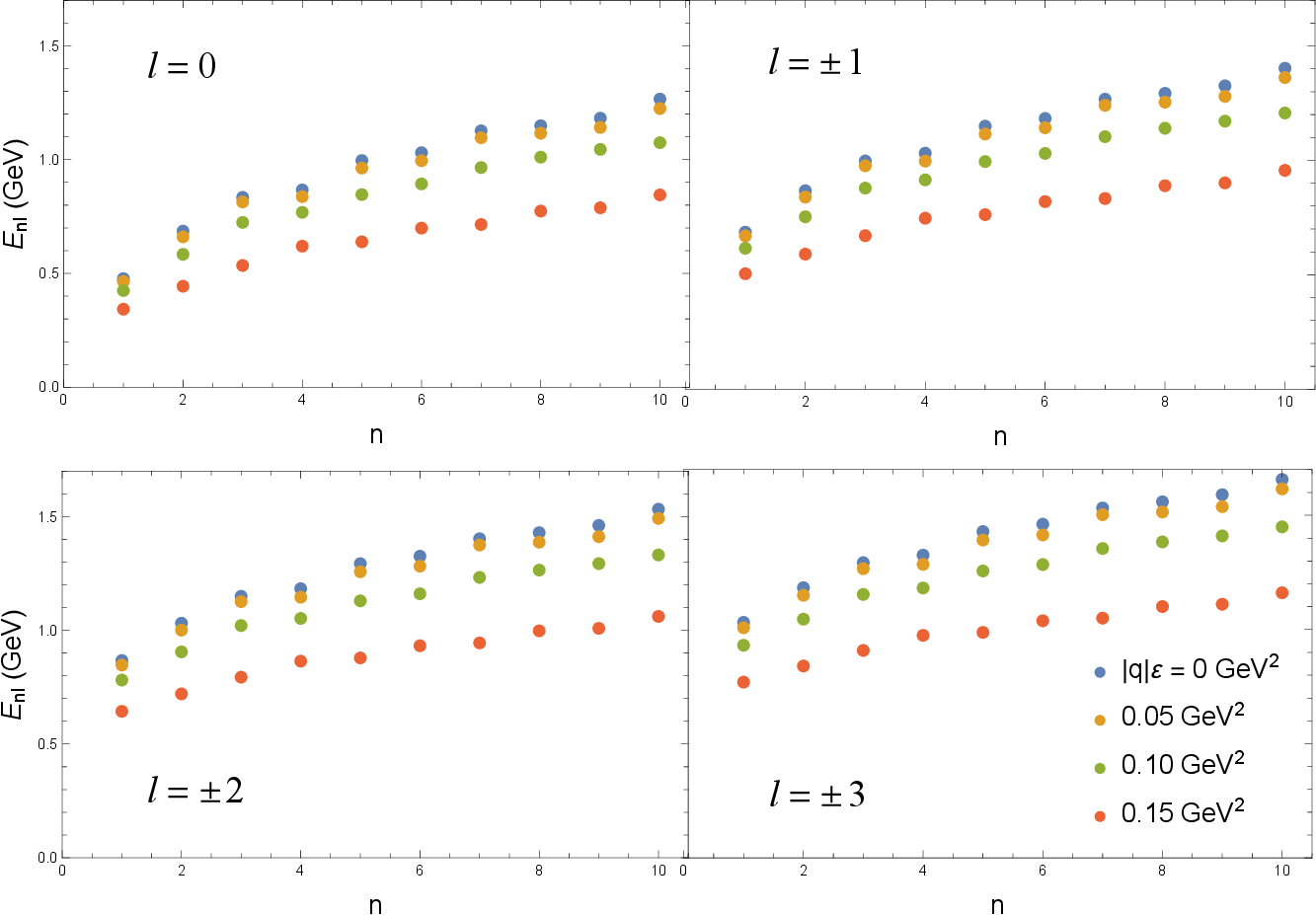}
	\caption{The lowest eigenenergies $E_{nl}$ of bottomium with $n=1,2,\dots,10$ for different subcritical electric fields and orbital angular momenta $l=0,\pm1,\pm2,\pm3$ along the electric field.}\label{Enn3}
	\end{center}
\end{figure}

\begin{figure}[!htb]
	\begin{center}
	\includegraphics[width=8cm]{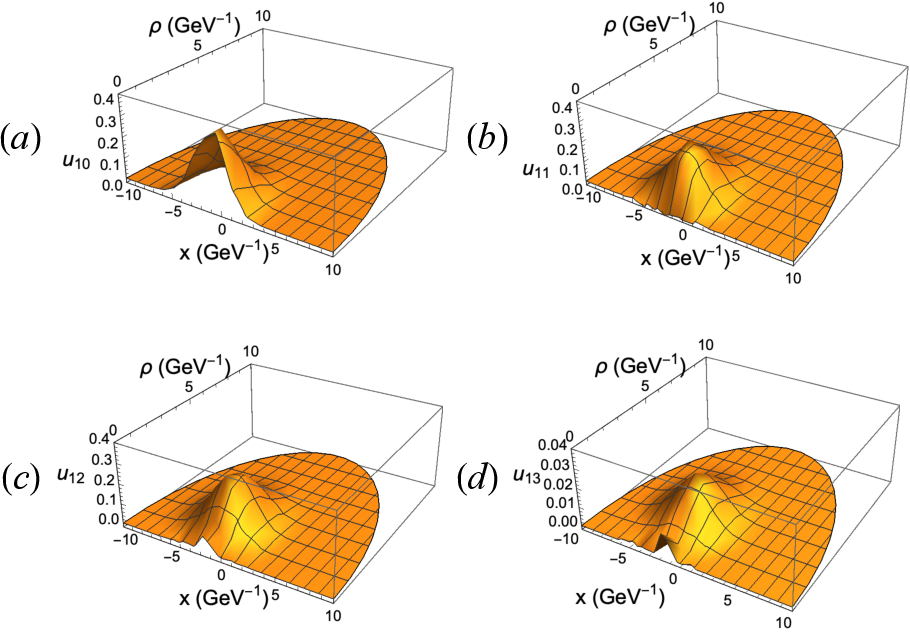}
	\caption{The lowest eigenfunctions $u_{nl}$ of bottomium for the subcritical electric field $|q|\varepsilon=0.1~{\rm GeV}^2$ and orbital angular momenta $l=0,1,2,3$ along the electric field. $(a), (b), (c)$ and $(d)$ correspond to $u_{10}(\rho,x), u_{11}(\rho,x), u_{12}(\rho,x)$ and $u_{13}(\rho,x)$, respectively. }\label{psinn3}
	\end{center}
\end{figure}

The full Schr$\ddot{\text o}$dinger equation \eqref{SE3C} can be solved numerically for different subcritical electric fields and the lowest $10$ eigenenergies $E_{nl}$ of bottomium are listed in Fig.~\ref{Enn3} for $l=0,\pm1,\pm2$ and $\pm3$. In the case $|q|\varepsilon=0$,  we find that the system is spherically symmetric and the eigenstates are $(2L+1)$-plet degenerate for a given total orbital angular momentum $L$ with $l=0,\pm1,\dots,\pm L$. Similar to those in Figs.~\ref{Enn1} and \ref{Enn2}, the eigenenergies get reduced by the external electric field, and the degeneracy becomes one or two since the system is only cylindrically symmetric for $|q|\varepsilon\neq0$. For demonstration, we show the lowest eigenfunctions for the electric field $|q|\varepsilon=0.1~{\rm GeV}^2$ and orbital angular momentum $l=0,1,2,3$ in Fig.~\ref{psinn3} : $(a), (b), (c)$ and $(d)$ correspond to $u_{10}(\rho,x), u_{11}(\rho,x), u_{12}(\rho,x)$ and $u_{13}(\rho,x)$, respectively. It is found that the centers of the wave functions shift antiparallel to the direction of the electric field, similar to those in one and two dimensions.

\begin{figure}[!htb]
	\begin{center}
	\includegraphics[width=8cm]{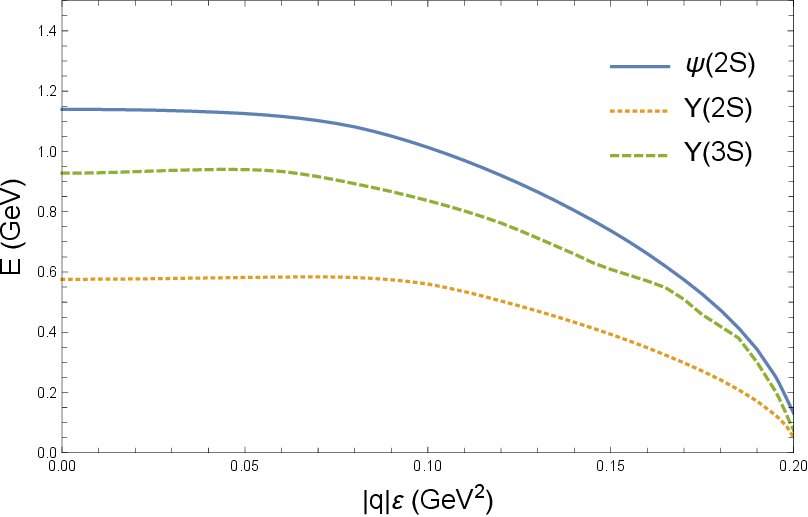}
	\caption{The relative eigenenergies of $\psi(2S), \Upsilon(2S)$ and $\Upsilon(3S)$-like mesons as functions of electric field $|q|\varepsilon$.}\label{PU}
	\end{center}
\end{figure}

Finally, to be more relevant to true QCD, we apply the study to charmonium and bottomium by adopting a realistic potential~\cite{Kawanai:2011jt}
\bea
V(r)=0.2\,r-{0.4105\over r}+{\beta}e^{-1.982r}{\bf{s_q\cdot s_{\bar{q}}}},
\eea
where ${\bf{s_q}}\ ({\bf{s_{\bar{q}}}})$ is the spin of quark (antiquark), $\beta=2.06$ for charmonium and $\beta=0.318$ for bottomium, and the units of $r$ and $V(r)$ are ${\rm GeV}^{-1}$ and ${\rm GeV}$, respectively. Since total angular momentum is no longer a good quantum number in an electric field, the well-known heavy-flavor mesons: $\psi(2S), \Upsilon(2S)$ and $\Upsilon(3S)$ are not eigenstates any more~\cite{Chen:2020xsr,Biaogang:2019hij}. But we will stick to the eigenstates initially corresponding to $\psi(2S), \Upsilon(2S)$ and $\Upsilon(3S)$ at vanishing $\varepsilon$, and numerical results are illustrated in Fig.~\ref{PU}. Though the relative energy monotonically decreases with $|q|\varepsilon$ for $\psi(2S)$-like meson, it slightly increases and then decreases with $|q|\varepsilon$ for $\Upsilon(2S)$ and $\Upsilon(3S)$-like mesonss. The increasing feature reflects the competition between effects of small electric field and Coulomb potential: Small $|q|\varepsilon$ enhances the average radii of the mesons and causes energy deficit, but the average Coulomb potential increases more with the radii. Around the critical point $|q|\varepsilon_c=\sigma$ of deconfinement, the relative energy decreases greatly for every quarkonium, but there is still a small positive redundant energy left at $|q|\varepsilon_c$ due to the spin-spin interactions. 

\section{Summary}\label{summary}
In this work, we fully explore the effect of electric field on heavy flavor mesons by considering one, two and three dimensions. For simplicity, we mainly focus on its interplay with the confining potential which dominates the internal interaction in mesons. In one dimension, the Schr$\ddot{\text o}$dinger equation can be solved analytically with the wave function given by Airy functions. When the strength of the electric field $|q|\varepsilon$ is smaller than the string tension $\sigma$, the wave function can automatically converge to zero for an infinite separation between the quark and antiquark. While, when $|q|\varepsilon>\sigma$, a boundary is necessary in the opposite direction of the electric field and the wave functions oscillate strongly near the boundary. In the viewpoint of quantum mechanics, the sudden changes of the features of the wave functions actually imply deconfinement transition of the heavy meson at $|q|\varepsilon_c=\sigma$. However, in the viewpoint of quantum field theory, $|q|\varepsilon_c=\sigma$ might only be the point where the initial heavy meson becomes unstable and could decay into two half-heavy mesons. For a larger electric field, even the half-heavy mesons cannot keep stable, so there should be a deconfinement transtion for the heavy meson. 

These qualitative discussions can be also applied to the cases with two and three dimensions, so only the unambiguous confined region $|q|\varepsilon<\sigma$ is considered thereafter. In two dimensions, the wave function can be presented as a product of elliptic sine/cosine functions and $r$-dependent functions in the large $r$ limit, and the quark-antiquark distribution is found to prefer antiparallel to the electric field. The eigenstates are singlet or doublet for $|q|\varepsilon=0$ but a finite electric field would break any degeneracy in two dimensions. In three dimensions, the wave function is similar to that in two dimensions in the large $r$ limit, so the quark-antiquark distribution still prefers antiparallel to the electric field. The electric field would break the degeneracy but not all -- there could still be double degeneracy left since the orbital angular momentum is a good quantum number along $\boldsymbol{\varepsilon}$. For all the dimensional cases with only confining potential, the electric field would reduce the eigenenergies and shift the centers of wave functions antiparallel to itself. However, if we recover the Coulomb potential and spin-spin interactions for more realistic studies, the relative energies of $\Upsilon(2S)$ and $\Upsilon(3S)$-like mesons would slightly increases with small $|q|\varepsilon$, though it monotonically decreases with $|q|\varepsilon$ for $\psi(2S)$-like meson.

\section*{Acknowledgment}
G. C. thanks Jiaxing Zhao for helpful comments. G. C. is supported by the Natural Science Foundation of Guangdong Province with Grant No. 2024A1515011225.

\end{document}